\documentclass[12pt]{article}
\usepackage{graphics}
\usepackage{amssymb,epsfig,amsmath,euscript,array}
\usepackage{axodraw4j}
\usepackage{subfigure}
\usepackage{color}
\usepackage{cite}

\makeatletter
\@addtoreset{equation}{section}
\makeatother

\definecolor{holger}{rgb}{0,0.5,0.7}
\definecolor{edit}{rgb}{1,0,0}

\definecolor{durbeer}{rgb}{1,0,0}

\definecolor{durbeer2}{rgb}{0.8,0,0.5}

\newcounter{multieqs}

\newcommand{\fssd}[1]{#1\!\!\!\!/}

\def\bd{\begin{document}}
\def\ed{\end{document}}
\def\nn{\nonumber}
\def\bea{\begin{eqnarray}}
\def\eea{\end{eqnarray}}
\let\bm=\bibitem
\let\la=\label

\begin{document}

\hfill{IPPP/09/30; DCPT/09/60}\\[-0.9cm]

\vspace{20pt}

\begin{center}

{\huge \bf  Probing Minicharged Particles\\[0.5ex] with Tests of Coulomb's Law} \\[1.5ex]

\vspace{30pt}

{\bf  Joerg Jaeckel}

{\small \em
{Institute for Particle Physics Phenomenology, Durham
University,\\ Durham DH1 3LE, United Kingdom}}

\vspace{10pt}

{\sffamily \tt
joerg.jaeckel@durham.ac.uk}

\vspace{30pt}
\end{center}

\begin{abstract}
Minicharged particles arise in many extensions of the Standard Model. Their contribution
to the vacuum polarization modifies Coulomb's law via the Uehling potential.
In this note we argue that tests for electromagnetic fifth forces can therefore be a sensitive probe
of minicharged particles. In the low mass range \mbox{$\lesssim \mu{\rm eV}$} existing
constraints from Cavendish type experiments provide the best model-independent bounds on minicharged particles.
\end{abstract}

\setcounter{page}{0}
\thispagestyle{empty}
\newpage


\section{Introduction}
Coulomb's inverse square law for the force between two charges is one of the central features of classical electrodynamics.
Its experimental discovery by Cavendish, Coulomb, and Robison was an important step in the development of
classical electrodynamics culminating in Maxwell's equations.
Over the last more than 200 years Coulomb's law has been tested with increasing precision and at
various length scales (s., e.g.,~\cite{Goldhaber:1971mr,Tu:2005ge}).
Indeed, the first detected deviation from it in the form of the Lamb shift was one of the first indications of ``new physics'' in the form
of \emph{quantum} electrodynamics~\cite{Lamb:1947,Bethe:1947,Lamb:1948}.

Nowadays, precise tests of Coulomb's law can serve as a powerful probe to search for new physics beyond the Standard Model.
A classical application is the search for a photon mass (s., e.g.,~\cite{Goldhaber:1971mr,Tu:2005ge}).
Beyond that, tests of Coulomb's law can even be used to search for new particles in so-called hidden sectors which interact only very weakly with
the particles of the Standard Model and which are accordingly difficult to detect in conventional collider experiments.
Due to their feeble interactions constraints on hidden-sector particles are often quite weak and they may be light with masses in
the eV-range or even below.
Hidden sectors appear in many extensions of the
standard model. In fact, it may be exactly those hidden sectors that give us crucial information
on how the standard model is embedded into a more fundamental theory as, e.g., string theory.
This makes it very desirable to have new, complementary probes of such particles.
One example of a hidden sector particle that can be searched for in using tests of Coulomb's law is massive extra U(1) gauge
bosons~\cite{Okun:1982xi,Popov:1999}.
In this note we will argue that they can also be used to search for another interesting class of hidden-sector
particles, namely particles with a small electric charge, so-called minicharged particles
(s.~\cite{Davidson:2000hf,Gluck:2007ia} for a test via the Lamb shift).

\begin{figure}[t]
\begin{center}
\includegraphics[angle=0,width=.6\textwidth]{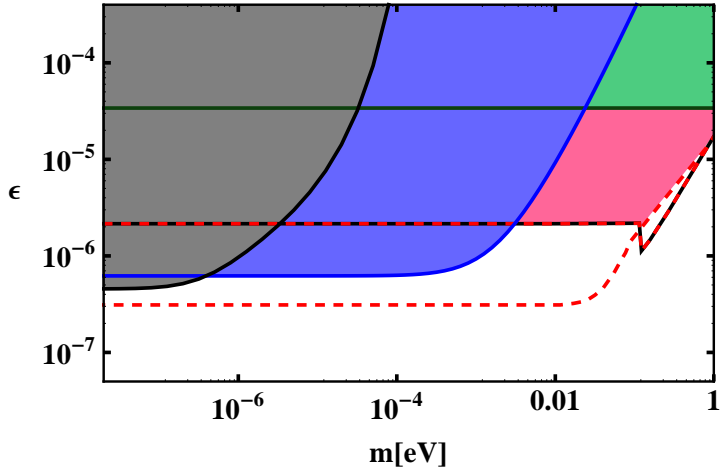}
\end{center}
\vspace{-0cm} \caption{Laboratory bounds on minicharged particles. The black line (on the left, bottom solid) corresponds to the exclusion limit obtained in this
note from the Cavendish type tests of Coulomb's law. The blue bound (on the left, middle solid) arises from constraints
on energy losses in high quality accelerator cavities~\cite{Gies:2006hv}.
The dark green curve (on the left, top solid) gives the limit arising from bounds on the invisible
decay on orthopositronium~\cite{Badertscher:2006fm,Ringwald} (a similar bound can be obtained from a reactor experiment~\cite{Gninenko:2006fi}).
The red-black dashed line denotes the limit~\cite{Ahlers:2007rd,Ahlers:2007qf} arising
from light-shining-through-a-wall experiments~\cite{Cameron:1993mr,Ehret:2007cm} and applies only to minicharged particles
arising from kinetic mixing$^3$ whereas the red dashed curve gives a limit~\cite{Gies:2006ca,Ahlers:2007qf}
from polarization experiments~\cite{Cameron:1993mr,Zavattini:2007ee}$^{4}$ and applies for a pure minicharged particle scenario. The shaded areas are
excluded in both scenarios.}
\label{bounds}
\end{figure}

Minicharged particles (MCPs) arise naturally and consistently\footnote{In theories with
kinetic mixing~\cite{Holdom:1985ag} minicharged particles are
indeed consistent with the existence of magnetic
monopoles~\cite{Brummer:2009cs}.} in a wide variety of extensions of the Standard Model based on field~\cite{Holdom:1985ag}
and string theory~\cite{Dienes:1996zr}.
Expectations for the size of their charge cover a wide range from $10^{-16}$ to $10^{-2}$~\cite{Holdom:1985ag,Dienes:1996zr}.
The best current laboratory limits in the sub-eV mass range are of the order of $\sim 10^{-6}$ (cf. Fig.~\ref{bounds}).
Astrophysical bounds (s., e.g.,~\cite{Davidson:2000hf}) are considerably stronger but also
somewhat model dependent~\cite{Masso:2005ym}\footnote{See, however,~\cite{Melchiorri:2007sq,Ahlers:2009kh} which combined give a less model
dependent bound~$\sim 10^{-9}$ from cosmology.}. Therefore, it is of particular importance to find new ways to search for minicharged particles in laboratory experiments.
In this note we will show that Cavendish type experiments searching for a deviation from Coulomb's law can be used for this purpose.

\newpage
\addtocounter{footnote}{1}
\footnotetext[\value{footnote}]{For a way to test pure minicharged particle models in a light-shining-through wall experiment see~\cite{Gies:2009wx}.}
\addtocounter{footnote}{1}
\footnotetext[\value{footnote}]{An interesting alternative to polarization experiments is interferometry~\cite{Dobrich:2009kd}.}

\section{Deviations from Coulomb's law from minicharged particles}
Minicharged particles cause a deviation from Coulomb's law via their effect on
the vacuum polarization in particular, via the Uehling potential~\cite{Uehling:1935}.
The presence of a non-vanishing vacuum polarization modifies the electric potential of a charge $Q$ (s., e.g.~\cite{Peskin:1995ev}),
\begin{equation}
\label{orig}
V(\mathbf{x})=V_{\rm Coulomb}(x)+\delta V(x)= Q\int\frac{d^{3}q}{(2\pi)^3}\exp(i\mathbf{q}\cdot\mathbf{x})\frac{e^2}{|\mathbf{q}|^2(1-\hat{\Pi}_{\epsilon}(q))},
\end{equation}
where $\hat{\Pi}_{\epsilon}(q)$ is an on-shell renormalized version of the vacuum polarization, e.g.,
\begin{equation}
\label{vacpol}
\hat{\Pi}_{\epsilon}(q^2)=-\alpha\epsilon^2\frac{2}{\pi}\int^{1}_{0}dx\,x(1-x)\log\left(\frac{m^2}{m^2-x(1-x)q^2}\right).
\end{equation}

For the vacuum polarization contribution of a particle of charge $\epsilon$ to the potential a standard textbook calculation~\cite{Peskin:1995ev} gives,
\begin{eqnarray}
\label{deviation}
\delta V(r)\!\!&=&\!\!
\frac{Q\alpha}{r}\left[\frac{2\alpha\epsilon^2}{3\pi}\int^{\infty}_{2m}dq\,\frac{\exp(-qr)}{q}\sqrt{1-\frac{4m^2}{q^2}}\left(1+\frac{2m^2}{q^2}\right)\right].
\end{eqnarray}
For large distances, $r\gg 1/m$, $\delta V$ drops off exponentially whereas for small distances, $r\ll1/m$, it's behavior is logarithmic as expected from the
running gauge coupling,
\begin{eqnarray}
\label{Uehlingapprox}
\delta V(r)\!\!&\approx&\!\! \frac{Q\alpha}{r} \left[\frac{\alpha\epsilon^2}{4\sqrt{\pi}}\frac{\exp(-2mr)}{(mr)^{\frac{3}{2}}}\right]
\quad\quad\quad\quad\quad\quad\quad\quad\quad\quad{\rm for}\quad mr\gg 1,
\\\nonumber
\!\!&\approx&\!\! \frac{Q\alpha}{r}\left[-\frac{2\alpha\epsilon^2}{3\pi}\log(2mr)-a\right],
\quad a\approx\frac{2\alpha\epsilon^2}{3\pi}\gamma \quad\quad\,\,\,\,{\rm for}\quad mr\ll 1.
\end{eqnarray}

So far we have considered the case where the minicharged particles have no further interactions. However, one of the most
natural ways in which minicharges arise is via kinetic mixing~\cite{Holdom:1985ag} of the electromagnetic U(1) gauge
boson with an extra ``hidden'' gauge boson under which the Standard Model particles are uncharged.
For such a theory the Lagrangian reads
\begin{equation}
{\mathcal{L}}=-\frac{1}{4} F^{\mu\nu}F_{\mu\nu}-\frac{1}{4}B^{\mu\nu}B_{\mu\nu}
-\frac{1}{2}\chi\,F^{\mu\nu}B_{\mu\nu},
\end{equation}
where $F_{\mu\nu}$ is the field strength tensor for the ordinary
electromagnetic U(1)$_{_\mathrm{QED}}$ gauge field $A^{\mu}$, and
$B^{\mu\nu}$ is the field strength for the hidden-sector
U(1)$_\mathrm{h}$ field $B^{\mu}$, {\it i.e.}, the hidden photon.  The first
two terms are the standard kinetic terms for the photon and hidden photon
fields. Because the field strength itself is gauge
invariant for U(1) gauge fields, the third term is also allowed by
gauge and Lorentz symmetry.  This term corresponds to a non-diagonal
kinetic term, a so-called kinetic mixing.
From the viewpoint of the low energy effective theory $\chi$ is a completely arbitrary parameter. Typical expectation in
extensions of the standard model range from $10^{-16}$ to $10^{-2}$~\cite{Holdom:1985ag,Dienes:1996zr}.
Accordingly we will treat $\chi$ in the following as a small parameter, $\chi\ll 1$.

The kinetic term can be diagonalized by a shift
\begin{equation}
\label{shift}
B^{\mu}\rightarrow \tilde{B}^{\mu}-\chi A^{\mu}.
\end{equation}
The only effect on the visible sector field $A^{\mu}$ is a multiplicative change of the gauge coupling,
\begin{equation}
\label{renorm}
e^2_{0}\rightarrow \frac{e^2_{0}}{1-\chi^2}\equiv e^2.
\end{equation}

We can now see how a minicharge arises for example for a  hidden matter fermion $h$
that has charge one under $B^{\mu}$. Applying the shift~\eqref{shift}
to the coupling term, we find:
\begin{equation}
e_h\bar{h}\fssd{B}\, h\rightarrow e_h\bar{h}\fssd{\tilde{B}}\, h-\chi e_h\bar{h}\fssd{A}\,
 h,
\end{equation}
where $e_h$ is the hidden-sector gauge coupling.
We can read off that the hidden-sector particle now has a small electric charge
\begin{equation}
\label{epsiloncharge}
\epsilon e=-\chi e_h\ll 1\quad\quad {\rm for}\quad \chi\ll 1
\end{equation}
under the visible electromagnetic gauge field $A^{\mu}$ which has
gauge coupling $e$.

The interesting questions is now what is the effect of a vacuum polarization caused by the hidden sector particle $h$, i.e. the minicharged particle.
The vacuum polarization can be most easily calculated in the unshifted basis where $h$ couples only to the hidden sector field $B^{\mu}$.
Accordingly its only effect is that the $B^{\mu}$ field has to be renormalized by a factor,
\begin{equation}
\sqrt{Z_{h}}=\sqrt{1-\hat{\Pi}_{h}(q^2)},
\end{equation}
where $\hat{\Pi}_{h}$ is given exactly by the same expression, Eq.~\eqref{vacpol}, as for $\hat{\Pi}_{\epsilon}$
only with $\alpha\epsilon^2$ replaced by the full hidden sector gauge coupling $\alpha_{h}$.
This renormalization entails that $\chi$, too, has to be renormalized,
\begin{equation}
\chi\rightarrow\frac{\chi}{\sqrt{Z_{h}}}.
\end{equation}
In turn, this then affects the renormalization Eq.~\eqref{renorm} of the ordinary electromagnetic gauge coupling after the diagnalization Eq.~\eqref{shift},
\begin{equation}
e^{2}_{0}\rightarrow\frac{e^{2}_{0}}{1-\frac{\chi^2}{Z_{h}}}=\frac{e^2_{0}}{1-\frac{\chi^2}{1-\hat{\Pi}_{h}(q^2)}}
\approx\frac{e^2_{0}}{(1-\chi^2)(1-\chi^2 \hat{\Pi}_{h}(q^2))}+{\mathcal{O}}(\chi^2 e^2\alpha^{2}_{h})
=\frac{e^2}{1-\hat{\Pi}_{\epsilon}(q^2)},
\end{equation}
where we have treated $\hat{\Pi}_{h}$ as a small parameter and have used the relations Eqs.~\eqref{renorm}, \eqref{epsiloncharge} to obtain the last equality.
The right hand side is exactly the expression that appears in Eq.~\eqref{orig}.
Accordingly, the vacuum polarization
effects in theories with kinetic mixing cause the same\footnote{Up to higher power in the hidden sector gauge coupling $\alpha_{h}$.}
deviations from Coulomb's law as in models with only minicharged particles.
The bounds derived in the following are therefore independent of the specific way in which the minicharge arises.

\section{Bounds from Cavendish type experiments}
Having identified the modification of the potential we can now turn to how it can be detected in experiments.
Currently, one of the most sensitive tests of the Coulomb potential is a variation of the concentric sphere setup used by Cavendish.
The idea behind this experiment is that if and only if the potential has a $1/r$ form the inside of a charged sphere is field free and
the potential accordingly constant.
In other words, the potential difference between a charged outer sphere and an uncharged inner sphere is zero if and only if the electric
potential has the Coulomb form. Deviations from this form would lead to a non-vanishing potential difference that can be measured.

In general, the potential of a sphere with a charge $Q$ and a radius $c$ at the distance $r$ from the center of the sphere is
\begin{equation}
U(Q,r,c)=\frac{Q}{2cr}\left[f(r+c)-f(|r-c|)\right],
\end{equation}
where
\begin{equation}
f(r)=\int^{r}_{0}ds\,s V(s,Q=1).
\end{equation}
It is straightforward to check that for a Coulomb potential $V_{\rm Coulomb}(r)=\alpha/r$, the potential
inside the sphere, i.e. for $r<c$, is constant. Explicitly one finds $f(r)=\alpha  r$ and $U(Q,r,c)=Q\alpha/c=const$.

In the simplest version of the Cavendish experiment one has an outer sphere of radius $b$, charged to a certain voltage, and then measures
the relative voltage difference to the uncharged inner sphere of radius $a<b$,
\begin{eqnarray}
\label{gamma1}
\gamma_{ab}=\left|\frac{{\mathcal V}_{b}-{\mathcal V}_{a}}{{\mathcal V}_{b}}\right|
\!\!\!&=&\!\!\! \left|\frac{U(Q,b,b)-U(Q,a,b)}{U(Q,b,b)}\right|
\\\nonumber
\!\!\!&=&\!\!\!\left|\frac{b}{\alpha Q}\frac{\delta U(Q,b,b)-\delta U(Q,a,b)}{1+\delta U(Q,b,b)b/(\alpha Q)}\right|
=\left|\frac{b}{\alpha}\frac{\delta U(1,b,b)-\delta U(1,a,b)}{1+\delta U(1,b,b)b/\alpha}\right|.
\end{eqnarray}
On the right hand it is understood that $\delta U$ is the part arising from the non-Coulomb part $\delta V$.
Using Eq.~\eqref{deviation} for $\delta V$ we can easily find,
\begin{equation}
\delta f(r)=\alpha^2\epsilon^2\frac{2}{3\pi}r\int^{\infty}_{2mr}dx\,\frac{1-\exp(-x)}{x^2}\sqrt{1-\frac{4(mr)^2}{x^2}}\left(1+\frac{2(mr)^2}{x^2}\right)
\equiv\alpha^2\epsilon^2 r\,g(mr).
\end{equation}
Inserting into \eqref{gamma1} we find,
\begin{equation}
\gamma_{ab}=\frac{\alpha\epsilon^2}{2}\frac{2 g(2bm)-(1+b/a) g((a+b)m)+(b/a-1) g((b-a)m)}{1+\alpha\epsilon^2g(2bm)}.
\end{equation}
We expect the best bounds for small masses $m\ll 1/a,1/b$. In this limit we have,
\begin{equation}
g(mr)\approx -\frac{2}{3\pi}\log(2mr),\quad{\rm for}\quad mr\ll 1.
\end{equation}
In this regime we therefore have,
\begin{eqnarray}
\nonumber
\gamma_{ab}\!\!&=&\!\!\frac{\alpha\epsilon^2}{3\pi}\left|\log\left(\frac{2b}{a+b}\right)+\log\left(\frac{2b}{b-a}\right)+\frac{b}{a}\log\left(\frac{b-a}{b+a}\right)\right|
+{\mathcal O}((\alpha\epsilon)^2)\quad\quad  ma,mb\ll 1
\\
\!\!&\sim&\!\! 0.05 \alpha\epsilon^2,
\end{eqnarray}
where we have assumed $b/a\sim 2$ in the second line.

In such a setup, Plimpton and Lawton~\cite{Plimpton:1936} achieved already 1936 a sensitivity of the order $|\gamma_{ab}|\lesssim 3\times 10^{-10}$.
This corresponds to a bound of the order of $\epsilon\lesssim 9\times 10^{-4}$.
Later experiments used a somewhat more complicated setup with several spheres. The latest and most precise~\cite{Williams:1971ms,Bartlett} uses four
spheres\footnote{To be precise they used icosahedrons but we will approximate those as spheres.} with radii $d>c>b>a$.
A very high voltage is applied between the outer two and the voltage difference is measured between the innermost pair.
One can easily derive the appropriate expression for this case,
\begin{eqnarray}
\label{gamma2}
\gamma_{abcd}=\left|\frac{{\mathcal V}_{b}-{\mathcal V}_{a}}{{\mathcal V}_{d}-{\mathcal V}_{c}}\right|
=\left|\frac{C\left[\delta U(b,c)-\delta U(a,c)-\delta U(b,d)+\delta U(a,d)\right]}{1+C\left[\delta U(c,c)+\delta U(d,d)-2\delta U(c,d)\right]}\right|,
\end{eqnarray}
where $C^{-1}=\alpha(1/c-1/d)$.
For this quantity the experiment achieves a precision~\cite{Williams:1971ms,Bartlett},
\begin{eqnarray}
\!\!\!\!\!\!\!\!\!\gamma_{abcd}\!\!&\lesssim&\!\! 2\times 10^{-16},
\quad\quad{\rm with}\quad\quad a=60\,{\rm cm},\,b=94\,{\rm cm},\,c=94.7\,{\rm cm},\,d=127\,{\rm cm}
\end{eqnarray}
From our earlier discussion we expect $\epsilon\lesssim {\rm few}\times 10^{-7}$ for small masses. For larger masses $m\gg 1/a$ we expect a rapid
weakening of the bound due to the exponential decay of the Uehling potential (cf. Eq.~\eqref{Uehlingapprox}) in this regime.
The bound obtained from a numerical evaluation of Eq.~\eqref{gamma2} is shown in Fig.~\ref{bounds} as the black area.
We can see that for small masses this is the best laboratory bound that applies for minicharged particles with and without an additional massless
hidden photon.

\section{Conclusions}
In this note we have shown that tests of Coulomb's law can be used to search for minicharged particles.
Minicharged particles leave their imprint on the potential between charges via their contribution to
the vacuum polarization resulting in the Uehling contribution to the potential.
Very precise laboratory limits on deviations from the Coulomb potential arise from Cavendish
type experiments. Using these we find a limit of $\epsilon\lesssim 5\times 10^{-7}$ in the mass range
below $0.1\,\mu{\rm eV}$  (cf. also Fig.~\ref{bounds}). In this mass range this is the best laboratory bound for minicharged particles in theories where
the minicharge arises from kinetic mixing.

As a final note we would like to point out that the Cavendish experiments~\cite{Williams:1971ms}
we used to derive our bounds were performed nearly 40 years ago. One could hope that with current technology
significant improvements are possible. Therefore, precision tests of Coulomb's law may yet again help us to explore the frontier of new physics.

\section*{Acknowledgements}
The author would like to thank Markus Ahlers, Holger Gies, Javier Redondo and Andreas Ringwald for helpful comments.


\end{document}